\documentclass[useAMS,usenatbib, usegraphicx, referee]{mn2e}

\usepackage{color}
\usepackage{ulem}

\def\Teff{$T_{\rm{eff}}$}
\def\Tcr{$T_{\rm{cr}}$}

\def\Tconst{$T_{\rm{const}}$}
\def\logg{log \textit{g}}

\def\HtO{$\mathrm{H_{2}O}$}
\def\Ht{$\mathrm{H_{2}}$}
\def\CHf{$\mathrm{CH_4}$}

\def\AKARI{\textit{AKARI}}

\def\mbf#1{\mbox{\boldmath ${#1}$}}

\title[A Signature of Chromospheric Activity in Brown Dwarfs]
{A Signature of Chromospheric Activity in Brown Dwarfs Revealed by 2.5--5.0~$\mu$m {\AKARI} Spectra}
\author[S. Sorahana et al.]
{S. Sorahana$^{1}$\thanks{E-mail:sorahana.satoko@j.mbox.nagoya-u.ac.jp} 
and T. K. Suzuki$^{1}$ and I. Yamamura$^{2}$\\
$^{1}$Department of Physics, Nagoya University, Nagoya, Aichi 464-8602, Japan\\
$^{2}$Department of Space Astronomy and Astrophysics, Institute of Space and Astronautical Science (ISAS),\\ 
Japan Aerospace Exploration Agency (JAXA), 
Sagamihara, Kanagawa 252-5210, Japan}

\begin{document}


\pagerange{\pageref{firstpage}--\pageref{lastpage}} \pubyear{2013}

\maketitle

\label{firstpage}

\begin{abstract}
We propose that the 2.7~$\mu$m {\HtO}, 3.3~$\mu$m {\CHf} and 4.6~$\mu$m CO absorption bands  
can be good tracers of chromospheric activity in brown dwarfs. 
In our previous study, we found that there are difficulties in  
explaining entire spectra between 1.0 and 5.0~$\mu$m 
with the Unified Cloudy Model (UCM), a brown dwarf atmosphere model. 
Based on simple radiative equilibrium, temperature in a model atmosphere usually decreases monotonically with height. 
However, if a brown dwarf has a chromosphere, as inferred by some observations, the temperature in the upper atmosphere is higher. 
We construct a simple model that takes into account heating due to chromospheric activity by setting a temperature floor in an upper atmosphere, 
and find that the model spectra of 3 brown dwarfs with moderate H{$\alpha$} emission, an indicator of chromospheric activity, are considerably improved to match the {\AKARI} spectra. 
Because of the higher temperatures in the upper atmospheres, 
the amount of {\CHf} molecules is reduced and the absorption band strengths become weaker. 
The strengths of the absorption bands of {\HtO} and CO also become weaker. 
On the other hand, other objects with weak H{$\alpha$} emission cannot be fitted by our treatment.  
We also briefly discuss magnetic heating processes which possibly operate in upper atmospheres, by extending our numerical simulations for the Sun and stars with surface convection to brown dwarf atmospheres.\end{abstract}


\section{Introduction}

\label{introduction}
Brown dwarfs are objects with mass intermediate between stars and planets.
No steady nuclear fusion takes place in their core, 
except for 
deuterium burning in the core of relatively massive and 
young ($\la10^6$~yr) brown dwarfs.
Hence, they simply cool off after the initial heating by gravitational energy / deuterium burning, 
and thermonuclear processes do not dominate their evolution (\citealt{Burrows_2001}). 
The first genuine brown dwarf, Gl~229B, was discovered by \citet{Nakajima_1995}, 
and studies of brown dwarfs are dramatically evolved in the last two dicades thanks to the development of instruments and models 
\citep{Tsuji_2002, Tsuji_2005, Allard_2001, Allard_2003, Ackerman_2001, Cooper_2003, Woitke_2003, Woitke_2004, Helling_2001, 2008MNRAS.391.1854H}.

The atmospheres of brown dwarfs are dominated by molecules and dust.
Many photometric/spectroscopic observations have been made in the near-infrared wavelength range 
shorter than 2.5~$\mu$m for studying the brown dwarf photosphere, 
because this wavelength range contains the spectral peaks of L dwarfs and is relatively easy to be observed. 
This wavelength range has features of various molecular species (e.g. TiO, VO, FeH, {\HtO} and {\CHf}) 
and effects of dust (e.g. Fe, Al$_2$O$_3$, MgSiO$_3$) extinction (\citealt{Burrows_2001, Tsuji_1996a, Tsuji_2002, Cushing_2006, 2008MNRAS.391.1854H}). 
Thus spectroscopic observations in the infrared regime are the most powerful tools to obtain physical 
and chemical information of brown dwarf photospheres. 
The radiation from inner photosphere becomes weaker by the dust extinction. 
The effect is different between spectral types, L and T.
Dust in the photosphere contributes to the spectra directly by dust extinction as well as indirectly 
by changing the structure of the photosphere. 
The effect of dust appears mainly at $J$ and $H$ bands in the spectra of L dwarfs (\citealt{Tsuji_1996b, Nakajima_2001}). 
Meanwhile, the spectra of T dwarfs are less affected by the dust opacity. 
This indicates that the dust settles lower in the photosphere of T dwarfs. 
In this manner, we understand the internal chemistry and physics 
with near-infrared spectral data shorter than 2.5~$\mu$m.  

However, many critical questions related to broader wavelength range 
spectra remain unanswered. 
\citet{Cushing_2008} reported that their model spectra result only poorly fits the observed spectra 
in the 0.95--14.5~$\mu$m for the mid- to late-L dwarfs and the early-T dwarfs. 
They used data observed by IRTF/SpeX (0.9--2.5~$\mu$m and 3.0--4.0~$\mu$m) 
and Spitzer/IRAC (5.0--14.5~$\mu$m). 
They concluded that the relatively poor fits at the L/T boundary, 
where dust contribution becomes smaller toward late type, 
are most likely due to the limitations of their simple cloud model \citep{Marley_2002}. 
In particular, the 3.0--4.0~$\mu$m range spectra resulted in the poorest fits. 
Observation in a wavelength range between 2.5 and 5.0~$\mu$m is difficult from the ground 
because of the Earth's atmospheric effects. 
Therefore, little spectral data has been obtained so far. 

{\AKARI}, a Japanese infrared astronomical satellite, obtained the spectral data of this wavelength range 
for 27 known brown dwarfs,
and we got 16 good quality data with ratio of signal to noise better than 3 (\citealt{Sorahana_2012}). 
They carried out the model fitting to each spectral data. 
They used shorter wavelength spectra (1.0--2.5~$\mu$m of IRTF/SpeX or UKIRT/CGS4; 
hereafter SpeX/CGS4\footnote{also see Section~\ref{spex} and \ref{CGS4} for detail}) 
supplementary in their analysis to derive the most probable physical parameter set (effective temperature, {\Teff}, 
surface gravity, {\logg}, and critical temperature, {\Tcr}, indicating the thickness of the dust layer)
in the model fitting. By using Unified Cloudy Model \citep[UCM hereafter;][]{Tsuji_2002,Tsuji_2005},
we search for the model atmosphere that simultaneously explains both the {\AKARI} and the SpeX/CGS4 spectra of each object reasonably well. 
However, we found that any combinations of the model parameters 
cannot give a reasonable fit to the observed data in the entire wavelength range 
(1.0--5.0~$\mu$m) of each object simultaneously, and  
any model spectra are always somewhat deviated from the observed spectrum in either the {\AKARI} or the SpeX/CGS4 wavelength.
The discrepancy implies that we are missing something important in the atmospheres of the brown dwarfs when constructing the model atmospheres.

In previous studies, X-ray, H${\alpha}$, and radio emissions, which indicate the presence of high temperature regions, from some brown dwarfs (\citealt{Stelzer_2006, Tsuboi_2003, Mohanty_2003, Schmidt_2007, Reiners_2008, Berger_2010, Hallinan_2007, Hallinan_2008}). 
Their observations show that the $L_{\rm{x}}/L_{\rm{bol}}$ ratio declines with {\Teff}, 
where $L_{\rm{x}}$ is X-ray luminosity and $L_{\rm{bol}}$ is bolometric luminosity.
However, relatively high X-ray luminosities $L_{\rm{x}}$ were observed in brown dwarfs whose spectral types are earlier than mid-L.  
In addition, H${\alpha}$ at 6563~{\AA} were observed (\citealt{Mohanty_2003, Schmidt_2007, Reiners_2008}).
The radio emissions from brown dwarfs were also detected (Hallinan et al., 2008). 
\citet{Kellett_2002} and \citet{Bingham_2001a} proposed that the origin of radio emissions may be electron cyclotron maser emission originating in the polar regions of a large-scale magnetic field.
From these observational results, the temperatures may increase somewhere in the upper atmospheres. In this paper, we call the heating region chromosphere, instead of photosphere whose temperature structure follows radiative equilibrium. Thus we need to reconsider the  thermal structures of brown dwarf atmospheres.

In this paper, we investigate how the broadband spectra of the observed brown dwarfs are affected by increases of the temperatures in the upper atmospheres assuming the existence of chromospheric and coronal activities.
We introduce the observational data of selected brown dwarfs in Section~\ref{Data}.
We carry out model fittings in Section~\ref{MF} without (\S \ref{nonheat}) 
and with (\S \ref{heat}) chromospheric heating.
Then, we discuss possible heating mechanism  
in Section~\ref{discussion}. 

\section{Observational Data} 
\label{Data}

\subsection{The {\AKARI} Sample}
\label{sample}
In this study, we focus on mid-L dwarfs from L4 to L7.5 types. 
The physics of early-L dwarfs may be different from brown dwarfs later than mid-L
because they are placed at the threshold of hydrogen burning. 
On the other hand, the chromospheric activity decreases toward later type dwarfs (\citealt{Gizis_2000, Mohanty_2003, Berger_2010}).
These authors statistically analyzed X-ray and H${\alpha}$ luminosities with spectral types, and concluded that the ratio of X-ray and H${\alpha}$ luminosities to the bolometric luminosity appears to decrease in the later spectral types.  
We therefore analyze the following mid-L dwarfs; 2MASS~J0036+1821 (L4), 2MASS~J2224--0158 (L4.5), GJ~1001B (L5), SDSS~J1446+0024 (L5), SDSS~J0539--0059 (L5), 2MASS~J1507--1627 (L5), 2MASS~J0825+2115 (L6) and 2MASS~J1632+1904 (L7.5). 
We summarize these objects in Table~\ref{radmasslist}. 
They are nearby and bright, thus generally well studied.

\begin{table*}
\begin{minipage}{126mm}
\caption{Eight Brown Dwarfs observed by {\AKARI} }\label{radmasslist}
\begin{tabular}{lcccc}
\hline
{Object Name}  &{Sp. Type} &  {Instrument} &  {References}  
               &  {cite of archive}\\
\hline
2MASS~J00361617+1821104 & L4 &SpeX&2&b\\
2MASS~J22244381--0158521 & L4.5 &SpeX&1&a\\
GJ~1001B & L5 &SpeX&1&b\\
SDSS~J144600.60+002452.0  &L5    &CGS4&2&c\\
SDSS~J053951.99--005902.0 &L5   &SpeX&2&b\\
2MASS~J15074769--1627386&L5  &SpeX&1&a\\
2MASS~J08251968+2115521 & L6 &SpeX&2&a\\
2MASS~J16322911+1904407 &L7.5  &SpeX&2&b\\
\hline
\end{tabular}
\medskip
Reference of spectral type (1) \citet{Kirkpatrick_2000}, 
(2) \citet{Geballe_2002}\\  
The data of (a) is given from the IRTF Spectral Library by Michael Cushing, 
that of (b) is from the SpeX Prism Spectral Libraries by Adam Burgasser, 
and (c) is the data given from Dagny Looper by private communication.
\end{minipage}
\end{table*}

\subsection{IRTF/SpeX Spectra}
\label{spex}
Almost all brown dwarfs except for SDSS~J1446+0024 in our sample of this analysis have been observed by \citet{Burgasser_2004,
Burgasser_2006, Burgasser_2008, Burgasser_2010, Burgasser_2007, Cushing_2004} with SpeX. 
SpeX is the medium-resolution 0.8--5.4~$\mu$m spectrograph mounted on the NASA Infrared Telescope Facility (IRTF), 
which is a 3.0 meter telescope at Mauna Kea, Hawaii.
The data have been obtained using its low-resolution prism-dispersed mode with the resolutions of 75--200, depending
on the used slit-width for three objects, 2MASS~J0036+1821, GJ1001B and 2MASS~J1632+1904. 
We retrieve these data from the SpeX Prism Spectral Libraries built by Adam Burgasser and Sandy Leggett\footnote{URL; http://pono.ucsd.edu/$\sim$adam/browndwarfs/spexprism/html/all.html}. 
Only SDSS~J0539--0059 spectrum was unpublished, and
we obtained from Mike Cushing (2010, private communication)\footnote{These data are now included in The SpeX Prism Spectral Libraries.}. Other three sources have
been observed by SpeX using its short wavelength cross-dispersed mode (SXD) with the resolutions
of 1200--2000, depending on the slit-width used. We get these data from the IRTF Spectral Library maintained by Michael Cushing\footnote{URL; http://irtfweb.ifa.hawaii.edu/$\sim$spex/IRTF\_Spectral\_Library/}.

\subsection{UKIRT/CGS4 Spectra}
\label{CGS4}
SDSS~J1446+0024 has not been observed with SpeX. 
A spectrum in 1.0--2.5~$\mu$m of SDSS~J1446+0024 was observed with UKIRT/CGS4 (\citealt{Geballe_2002}). 
CGS4 is the multi-purpose grating spectrometer equipped on the 3.8~m United Kingdom Infrared Telescope(UKIRT), 
which is also sited on Hawaii Mauna Kea. 
CGS4 has four gratings. 
The data for SDSS~J1446+0024 was observed using the 40 line/mm grating that provided the resolution of 300--2000 or $400\times\lambda$ $\mu$m. 
The spectrum was taken by adopting two broad band filters for the low and the medium resolution gratings in use with CGS4, 
namely B1 and B2, and the wavelength range of these filters are 1.03--1.34~$\mu$m 
and 1.43--2.53~$\mu$m, respectively. 
We get the spectral data of SDSS~J1446+0024 from Dagny Looper (2010, private communication). 

\section{Model Fitting}
\label{MF}
In our previous study, we searched for the model atmospheres that explain both the {\AKARI} 
and the SpeX/CGS4 spectra of the brown dwarfs reasonably well (\citealt{Sorahana_2012}).
While the wavelength range of {\AKARI} reflects the condition of relatively upper atmospheres
(Sorahana et al. 2013 in preperation), 
that of SpeX/CGS4 is sensitive to the inner atmosphere including the effect of dust lying in the inner atmospheres
(where $\tau\sim1$). 

In this paper, we take a different fitting strategy from \citet{Sorahana_2012} to investigate the temperature structures of the upper atmospheres affected by the presence of chromospheres. 
In order to pin down the thermal structures in the inner atmospheres, we first carry out the model fittings to the only SpeX/CGS4 spectral data (\S \ref{fitting}). 
We call the model atmospheres determined in this way ``non-heating best-fit models''. 
As shown in \S \ref{fitresult}, none of the non-heating best-fit models shows perfect fit to the observed spectrum in the entire wavelength range. 
As the second step, we modify the temperature structures in the upper atmospheres assuming the presence of chromospheres/coronae and seek model atmospheres that give better fits to the observations (\S \ref{revisedmodel}). 
We call the model atmospheres derived by the second step ``heating best-fit models''. 
In Figures~\ref{fig1} and \ref{fig2} we display the spectra synthesized from the non-heating best-fit models ({\it green} lines) 
and those from the heating best-fit models ({\it red} lines)\footnote{Green lines of Figure~\ref{fig2} are not heating best-fit model, but heating model with $f_{\rm{const}}$ of 0.8 (see Section~\ref{revisedmodel} for detail about $f_{\rm{const}}$).} in comparison with the observed spectra ({\it black} lines). 
We classify the eight brown dwarfs into two groups: In the first group consisting of three objects (Figure \ref{fig1}) 
the heating model spectra give reasonable fits to the observed spectra, while in 
the second group consisting of the others (Figure \ref{fig2}) the heating models still give poor fits to the observations (\S \ref{result}).

\subsection{Non Heating Models} 
\label{nonheat}
\subsubsection{Fitting Procedure} 
\label{fitting}
We derive physical parameters of the {\AKARI} objects, namely effective temperature {\Teff}, 
surface gravity {\logg} and critical temperature {\Tcr} by model fitting to the only SpeX/CGS4 spectral data with Unified Cloudy Model 
(UCM; \citealt{Tsuji_2002,Tsuji_2005}). 
{\Tcr} is given as an additional parameter in UCM that controls the dust dissipation thus the thickness of the dust layer. 
The UCM applies a simple concept with phase-equilibrium \citep{Tsuji_1996b}, and does not include the detail of cloud formation and growth mechanisms associated with hydrodynamic processes (\citealt{Woitke_2003, Woitke_2004, Helling_2006, 2008A&A...485..547H}). {\Tcr} cannot be determined from the physical theory but must be determined from observations empirically.
For {\Tcr}$<T<T_{\rm{cond}}$, dust condensation and sublimation are balanced.  
This means that the dust would exist only in the layer of {\Tcr}$<T < T_{\rm{cond}}$.

We follow \citet{Cushing_2008} and evaluate the goodness of the model fitting to the only shorter wavelength spectra by the statistic $G_k$ defined as
\begin{equation}
\label{gk}
G_{k} = \frac{1}{n-m}\sum_{i=1}^n \omega_{i}  \left( \frac{f_{i} - C_{k}F_{k,i}}{\sigma_{i}} \right)^2,
\end{equation}
where $n$ is the number of data points; $m$ is degree of freedom (this case $m=3$); $\omega_{i}$ is 
the weight for the $i$-th wavelength points 
(we give the equal weight as $\omega_{i}$ = 1 for all data points because of no bias within each observed spectrum); 
$f_{i}$ and $F_{k, i}$ are the flux densities of the observed data and $k$-th model, respectively; 
$\sigma_{i}$ are the errors in the observed flux densities 
and $C_{k}$ is the scaling factor given by 

\begin{equation}
\label{scalingfactor}
C_{k} = \frac{\sum \omega_{i} f_{i} F_{k,i}/\sigma_{i}^2}{\sum \omega_{i} {F_{k,i}}^2/{\sigma_{i}}^2}.
\end{equation}
$G_{k}$ is equivalent to reduced $\chi ^{2}$, since we adopt $\omega_{i}$ = 1 in our analysis. 
This method is same with that in \citet{Sorahana_2012}, except for fitting wavelength range. 

\subsubsection{Results} 
\label{fitresult}
We show the model spectra of the non-heating best-fit models ({\it green} lines), 
which use the only SpeX/CGS4 data for the fittings, in Figures ~\ref{fig1} 
\& \ref{fig2}.
We see that the non-heating best-fit models well explain the SpeX/CGS4 spectra, 
but the model spectra do not match with the observations in the {\AKARI} wavelength range well. 
The principal differences between the observed and model spectra from the non-heating best-fit models are  
seen in the flux levels in the {\CHf} at the 3.3~$\mu$m band and around the 4.0~$\mu$m region.
For instance, the {\CHf} bands of the three brown dwarfs, 2MASS~J2224--0158, GJ~1001B and 2MASS~1632+1904, cannot be explained. 
The model spectrum of GJ~1001B contradictorily exhibits the {\CHf} absorption feature at the 3.3~$\mu$m band, 
whereas it can reprocuce the overall observed spectrum from 1.0 to 5.0~$\mu$m.  
There are also differences at 2.7~$\mu$m {\HtO} and 4.6~$\mu$m CO bands in the spectrum of 2MASS~J2224--0158.
For other four objects, SDSS~J1446+0024, SDSS~J0539--0059, 2MASS~J1507--1627 and 2MASS~J0825+2115, 
the entire {\AKARI} spectra cannot be explained by the non-heating best-fit models especially the flux levels around 4.0~$\mu$m. 
The deviation of 2MASS~J0036+1821 shows a different trend from 
other objects; 
i.e., the flux level around 4.0~$\mu$m in the observed spectrum is lower than that in the model spectrum, 
even if 2.7~$\mu$m {\HtO} and 4.6~$\mu$m CO bands reasonably fit well to the observation. 
These results indicate that the SpeX/CGS4 data cannot solely constrain the physical parameters of the upper atmospheres of these observed brown dwarfs. 

\begin{table*}
\begin{minipage}{150mm}
\caption{Physical Parameters of Eight Brown Dwarfs observed by {\AKARI} }\label{table2}
\begin{tabular}{lcrrrrrr}
\hline
{Object Name}  &{Sp. Type} &{{\Tcr}[K]}&  {{\logg}}  &  {{\Teff}[K]}  
               & {best $T_{\rm{const}}$[K]} & {best $T_{\rm{const}}/${\Teff}} 
               & {log$(L_{\rm{H\alpha}}/L_{\rm{bol}})$}\\
\hline
2MASS~J00361617+1821104 & L4 &1800&5.5&1900&n/a&n/a&-6.26$^{\rm a}$\\
2MASS~J22244381--0158521 & L4.5 &1800&5.0&1700&1445&0.85&-6.48$^{\rm a}$\\
GJ~1001B & L5 &1800&5.0&1800&1260&0.70&-7.42$^{\rm a}$\\
&&&&&&&-5.23$^{\rm b}$\\
SDSS~J144600.60+002452.0  &L5    &1700&5.0&1800&n/a&n/a&no data\\
SDSS~J053951.99--005902.0 &L5   &1800&5.5&1900&n/a&n/a&no data\\
2MASS~J15074769--1627386&L5  &1800&5.5&1900&n/a&n/a&-8.18$^{\rm a}$\\
2MASS~J08251968+2115521 & L6 &1800&5.0&1700&n/a&n/a&-8.18$^{\rm a}$\\
2MASS~J16322911+1904407 &L7.5  &1800&5.5&1600&1280&0.80&-6.23$^{\rm b}$\\
\hline
\end{tabular}
\medskip
Reference of {log$(L_{\rm{H\alpha}}/L_{\rm{bol}})$}\\ 
a: \citet{Reiners_2008}, b: \citet{Mohanty_2003}. 
\end{minipage}
\end{table*}

\subsection{Heating Models} 
\label{heat}
\subsubsection{Revising Thermal Structure}
\label{revisedmodel}
The temperature calculated from UCM as well as other models assuming the radiative equilibrium that decreases monotonically with an increasing altitude.  
On the other hand, some brown dwarfs exhibit activities regarding chromospheres, coronae, and flares, as discussed in \S \ref{introduction}.   
In such objects, the temperatures eventually stop decreasing and turn to increase somewhere in the upper atmospheres. 
There are several possibilities to account for the temperature inversion, which is discussed later in \S \ref{discussion}. 
In this section, leaving the detailed heating mechanisms aside, we adopt a very simple procedure to take into account the effect of the modified temperature structures.
Since the temperature structures of the non-heating best-fit models are derived mainly from the $J$ and $H$ band features, 
which are sensitive to the effect of dust, we can reasonably assume that the temperature structures in the dust layers located in the inner photospheres are reliable even in the non-heating best-fit models. 
Thus we change the temperature structures only above the dust layers. 
We put a floor value, $T_{\rm const}$, for the temperature structure of 
each object in the following way: 
\begin{equation}
\label{eq1}
T(r) = \max(T(r),T_{\rm const})=\max(T(r),f_{\rm const}T_{\rm eff}), 
\end{equation}
where $f_{\rm{const}}$ is a parameter which is tuned by comparing 
the observed spectrum of each object(\S \ref{result}).
In other words, the surface temperature structure is replaced with a constant value following equation~\ref{eq1}, 
instead of that based on the radiative equilibrium (see also middle panel of Figure~\ref{fig3} for example).
The gas pressure remains unchanged to keep the hydrostatic equilibrium by reducing the density compared with the case without {\Tconst}, following the equation of state for an ideal gas, $p=(\rho/\mu m)kT$, where $p$, $\rho$, $\mu$, $m$, $k$, and $T$ are pressure, density, mean molecular weight, atomic mass unit, Boltzmann constant, and temperature, respectively.
We do not take into account the inversion of the temperatures but 
see how the model spectra are modified when the temperatures do not 
decrease in the upper atmosphere. 
In a sense this is a minimal requirement to consider a chromosphere and/or corona.
Using this heating model atmosphere, we solve the chemical equilibrium and then calculate the radiative transfer.

\subsubsection{Results} 
\label{result}
We explore how the inclusion of $T_{\rm const}$ improves the model spectra. 
By varying $f_{\rm const}$, we seek for the heating best-fit model for each object.
We show the photosphere structure in Figure~\ref{fig3} for the model of ({\Tcr}/{\logg}/{\Teff}) = ($1800K/5.0/1700K$) 
corresponding to the atmosphere of 2MASS~J2224--0158. 
The top panel of this figure shows the spectra of best-fit models without ({\it green}) and with ({\it red}) heating between 1.0 and 5.0~$\mu$m. 
The middle panel shows the temperature structures of these 
model atmospheres as a function of total gas pressure. 
The bottom panel shows the partial pressures of each molecule versus total gas pressure. 
We can see that the spectral shape in the range shorter than 2.5~$\mu$m, including $J$, $H$ and $K$ bands, does not change significantly, but that of {\AKARI} wavelength range, 2.5 to 5.0~$\mu$m, changes appreciably. 
From the bottom panel, we find that the {\CHf} abundance  
in the upper region changes dramatically by introducing $T_{\rm const}$.
This fact is reflected in the spectral feature around 3.3~$\mu$m shown in the top panel; i.e.,  
the absorption feature of the 3.3~$\mu$m {\CHf} band is diminishing.  
In addition, the absorption bands of 2.7~$\mu$m {\HtO} and 4.6~$\mu$m
CO in the heating model spectra tend to become weak.
In general, the strengths of the absorption bands are a result of
radiative transfer in which many factors such as number densities of
molecules, excitation, velocity structure, and relation to the
continuum source. Hence it is often difficult to identify
a unique reason for the variation.
In the current case, the higher temperature in the upper photosphere
cancels the effects of increased abundance of the molecules and make
the absorption even weaker.

\begin{figure}
\begin{center}
\includegraphics[width=90mm]{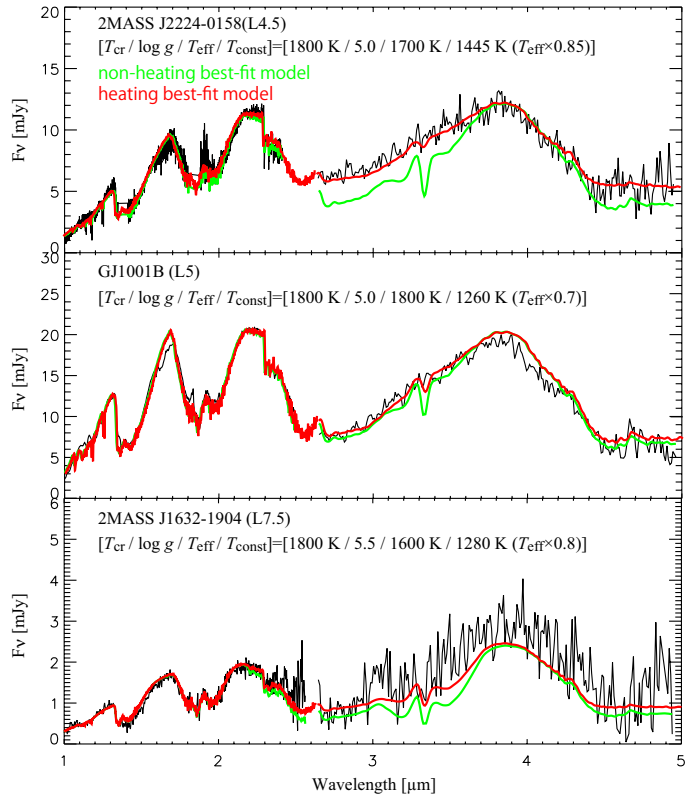}
\end{center}
\caption{Comparison of the model spectra with the observed spectra 
for the three objects, 2MASS~J2224--0158, GJ~1001B and 2MASS~1632+1904, which 
are well explained by the heating model atmospheres taking into account 
the heating in the upper atmospheres.
The {\it black}, {\it green}, and {\it red} lines respectively correspond to the observed spectra, the non-heating best-fit model spectra, and the heating best-fit model spectra.}
\label{fig1}
\end{figure}

We find that three of the eight objects, 2MASS~J2224--0158, GJ~1001B and 2MASS~1632+1904, 
can be explained by this new treatment, but the others cannot be improved sufficiently. 
For the three successful objects, 
the best-fit values of $T_{\rm{const}}$ are 1445 K ($T_{\rm{eff}} \times 0.85$), 1440 K ($T_{\rm{eff}} \times 0.70$), 
and 1280 K ($T_{\rm{eff}} \times 0.80$) for 2MASS~J2224--0158, GJ~1001B, and 2MASS~J1632+1904, respectively. 
We show the results in Figure~\ref{fig1} and Table~\ref{table2}. 
In the case of 2MASS~J2224--0158, the non-heating best fit model cannot reproduce the {\AKARI} spectrum, except for 3.8 to 4.3~$\mu$m. 
On the other hand, the heating best-fit model of 2MASS~J2224--0158 can explain the entire observation perfectly within the error. 
For GJ~1001B , there is small deviation between the heating and non-heating best-fit model spectra, except for {\CHf} absorption band at 3.3~$\mu$m (also see Section~\ref{fitresult}). 
If we consider the additional heating at upper atmosphere, the {\CHf} band strength fits better to observation. 
Although 2MASS J1632-1904 has less S/N than the other two objects, we see that its entire spectra of the heating model fits to the observation better than that of the non-heating model.

\begin{figure}
\begin{center}
\includegraphics[width=90mm]{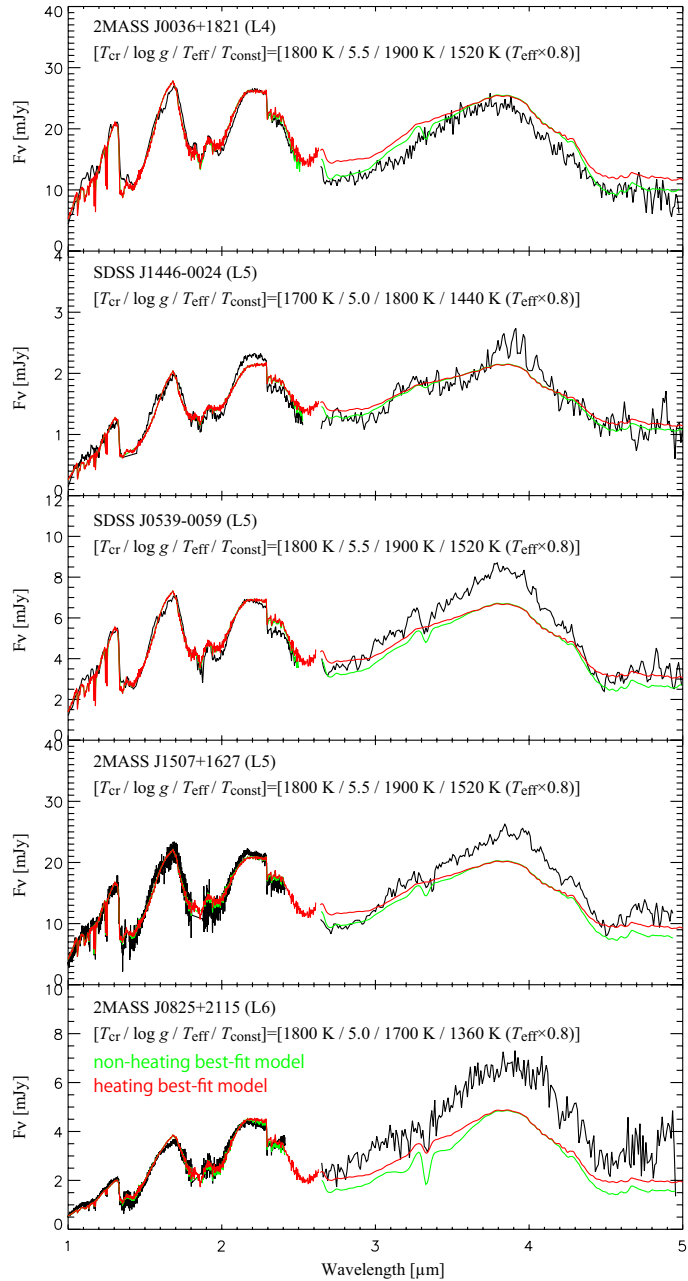}
\end{center}
\caption{Same as Figure \ref{fig1} but for the rest of the five brown 
dwarfs, 2MASS~J0036+1821, SDSS~J1446+0024, SDSS~J0539--0059, 2MASS~J1507--1627, and 2MASS~J0825+2115, which cannot be well fitted by the heating model spectra.
}
\label{fig2}
\end{figure}

\begin{figure}
\begin{center}
\includegraphics[width=95mm]{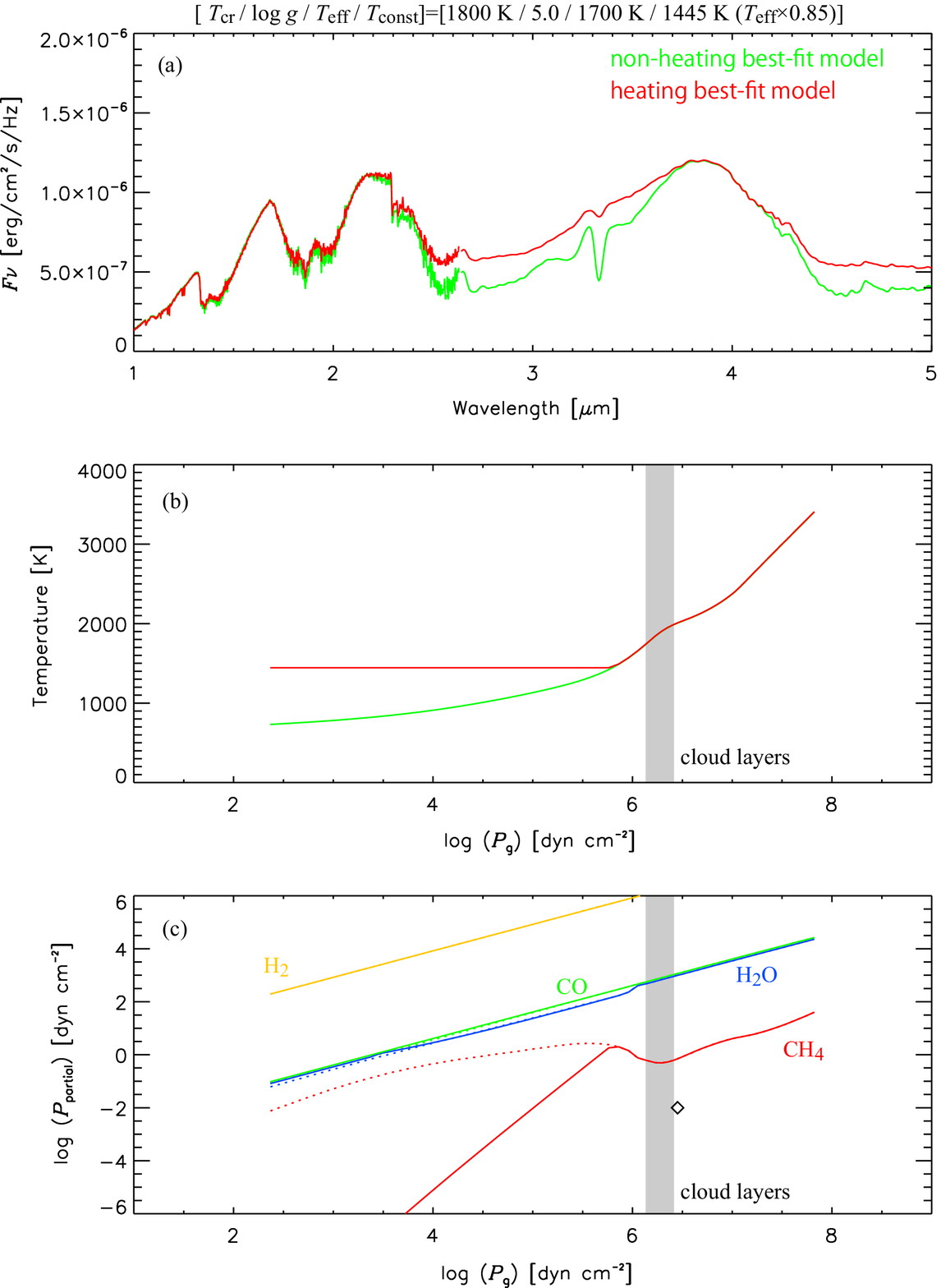}
\end{center}
\caption{Comparison of spectrum, temperature and chemical structure of the L dwarf model 
({\Tcr}/{\logg}/{\Teff}) = (1800K/5.0/1700K) for being constant lower than 1445 K ({\Teff}$\times0.85$). 
(a) The spectra of the models with ({\it red}) and without ({\it green}) heating. 
(b) The variation of temperature from that of the non-heating model. Colors are same with panel (a). 
(c)  Total pressure log $P_g$ versus partial pressures of {\Ht} ($\sim$ total log$P_g$), CO, {\HtO} and {\CHf}, 
which become dust, molecules, which are drawn with yellow, green, blue and red, respectively. 
The values of the non-heating model are drawn with dashed lines, and that of the heating model are drawn with solid lines.
Grey region shows dust layers.
}
\label{fig3}
\end{figure}

Figure~\ref{fig2} shows an example of the comparison between the observations and the heating models with 
$f_{\rm{const}}$ of 0.8 for the other five objects. 
As shown in this figure, it is seen that the heating model spectra ({\it red}) fit better than the non-heating model spectra ({\it green}) especially for 4.6~$\mu$m CO band, 
except for 2MASS~J0036+1821.
However, the flux levels around 4.0~$\mu$m do not improve even in the heating model spectra.
Thus, these objects with a deviation around 4.0~$\mu$m between the observation and the non-heating model cannot be explained by the modified temperature 
structure.

The trend of change for the revised model spectra for any stellar parameter for mid-L brown dwarfs is almost the same; i.e., only spectral features around 3.3~$\mu$m, 2.7~$\mu$m, and 4.6~$\mu$m change. In other words, the spectra shorter than 2.5~$\mu$m do not change. This is because wavelengths shorter than 2.5~$\mu$m are sensitive to the relatively inner photosphere which we do not change at all in our current analysis. As shown in \citet{Cushing_2008}, model fitting using narrow wavelength range spectra provide better fits than using wide wavelength range spectra at the same time. \citet{Sorahana_2012} analysed wide wavelength range spectra from 1.0 to 5.0~$\mu$m for model fitting to derive the most probable physical parameters for each object. They found that there are always some deviations between the observed and model spectra. For example, 2MASS J2224--0158 (L4.5), which is explained completely by our current heating model, has a large deviation at the $K$ band, which is located in wavelength range shorter than 2.5~$\mu$m. Thus, when we start with the stellar parameters derived in \citet{Sorahana_2012}, the heating model spectrum cannot reproduce the observed spectrum.

\section{Discussion}
\label{discussion}
As shown in Figure~\ref{fig1}, the model spectra of
the three mid-L dwarfs 
are considerably improved to match the observed spectra. 
These models take into account the temperature floors, $T_{\rm const}$, 
in the upper atmosphere.
However, the other five objects cannot be well fitted only by including $T_{\rm const}$ (Figure~\ref{fig2}).
The motivation to introduce $T_{\rm const}$ is to minimally take into account the effect of the heating in the upper atmosphere concerning chromospheric and coronal activity.
Thus it is considered that the three successful objects may have chromospheric and/or coronal activities, and the other objects do not have such strong activities.
Each object could potentially be in a different state of enhanced activity, e.g., a flare, or have different effective temperatures or different ages. For example, \citet{Berger_2010} discussed that X-ray luminosity decreases towards later spectral types. They suggest that this trend is caused by the dissipation of magnetic field at later spectral types.
We discuss in the following section firstly from an observational side and then from a theoretical side.

\subsection{Relation with Chromospheric Activities}
X-rays (\citealt{Stelzer_2006,Tsuboi_2003}) and H${\alpha}$ emissions (\citealt{Mohanty_2003, Schmidt_2007, Reiners_2008}) are detected in some brown dwarfs. 
These observations suggest that at least some brown dwarfs have hot regions implicating chromospheres and/or hot coronae in their upper atmospheres. 
We investigate the relation between our result and observed H${\alpha}$ emissions, which can be used as an indicator of chromospheric activity, of several brown dwarfs. 

In Table \ref{table2} we list H${\alpha}$ emission normalized by the bolometric luminosity, $L_{\rm{H\alpha}}/L_{\rm{bol}}$, 
by \citet{Mohanty_2003} and \citet{Reiners_2008}; see also \citet{McLean_2012} who compiled some H${\alpha}$ observations including \citet{Reiners_2008}. 
Among the eight objects, two are available in \citet{Mohanty_2003}  
and five are included in \citet{Reiners_2008}. 
GJ~1001B is in the both papers.
No data is available for the rest of two objects.  
The values for GJ~1001B in the two papers are different by two orders of magnitude. 
A possible explanation is that this object is very active and exhibits large time-variability related to flares. 
While the $L_{\rm{H\alpha}}/L_{\rm{bol}}$ generally decreases toward later type objects (\citealt{Reiners_2008}), 
the latest one (2MASS~1632+1904) among the eight shows rather large $L_{\rm{H\alpha}}/L_{\rm{bol}}$, which might be caused by high time-variability.

Among the six objects with $L_{\rm{H\alpha}}/L_{\rm{bol}}$, 
we first discuss the five objects except for 2MASS~J0036+1821.
The three objects, 2MASS~J2224--0158, GJ~1001B and 2MASS~J1632+1904, are inferred to possess high chromospheric activity from
their relatively large $L_{\rm{H\alpha}}/L_{\rm{bol}}$. 
Interestingly enough, they are the objects whose spectra are well reproduced by the heating models.
On the other hand, other two objects, 2MASS~J1507--1627 and 2MASS~J0825+2115, which have much lower $L_{\rm{H\alpha}}/L_{\rm{bol}}$, cannot be explained even though the temperature floors are considered. 
We should consider alternative effect 
for these unsuccessful objects.

The final one of the six objects, 2MASS~J0036+1821, with $L_{\rm{H\alpha}}/L_{\rm{bol}}$ appears to be an outlier. 
The deviation of the non-heating best-fit model spectrum from the observed spectrum appeared in 2.5 to 5.0~$\mu$m  
is the opposite direction from the other objects;  
the flux level of the model spectrum is higher than that of the {\AKARI} observed spectrum. 
Apart from the absolute magnitude flux level, 
the spectral shape in the {\AKARI} wavelength range itself seems to be improved, 
which might imply that our revised model partly makes sense in some respects in this object.

\subsection{Magnetic Heating} 
We conclude that the additional heating in an upper atmosphere is important to understand observed spectra of brown dwarfs. 
So far we have not specified mechanisms that account for the heating 
to keep the temperatures in the upper atmospheres. The surface region of a brown dwarf is convectively unstable, 
and it is considered that the energy is upwardly transported by the convection 
\citep{Baraffe_2002, Mohanty_2007}. 
We expect that magnetic fields are generated by dynamo actions, 
similarly to what takes place in the Sun and stars with a surface convective layer \citep[e.g.,][]{Choudhuri_1995,Hotta_2012}. 
Various types of magnetic waves are generated and a fraction of them propagates upwardly to heat up upper regions of the atmospheres and drive the stellar winds. 
The Alfv\'{e}n wave, among others, is a promising candidate that transfers the energy of the convection to upper regions, 
and leads to various magnetic activities such as chromospheres, coronae, and stellar winds, under the conditions of the Sun and other stars with surface convection. 
One of the authors of the present paper has studied various objects with surface convection, including the Sun \citep{Suzuki_2005,Suzuki_2006,Matsumoto_2012}, red giants \citep{Suzuki_2007}, active solar-type stars \citep{Suzuki_2013}, and hot jupiters \citep{Tanaka_2013}. 
Surface convection triggers the processes introduced here. 
Since brown dwarfs posses a surface convective layer, the similar processes 
could operate in their atmospheres.
Here, we demonstrate how the temperature structure of the model of ({\Tcr}/{\logg}/{\Teff}) = (1800~K/5.0/1700~K) corresponding to the non-heating best-fit model for 2MASS~J2224--0158 is affected by magnetic heating with a MHD (magnetohydrodynamical) simulation. 

We use the same simulation code originally developed for the Sun \citep[see][for the details]{Suzuki_2005, Suzuki_2006,Suzuki_2013}. 
We dynamically solve the structure of the atmosphere without assuming hydrostatic equilibrium. 
The temperature and density (accordingly gas pressure) dynamically change with time by the propagation and dissipation of waves; 
since Alfv\'{e}n waves accompany Poynting flux, their dissipation leads to the heating of the ambient gas. 
We do not solve radiative transfer but use a simplified radiation cooling rate empirically determined from observation of the solar chromosphere 
\citep{Anderson_1989}. We also adopt the ideal MHD approximation; we assume 
that the magnetic field is well-coupled with the gas component. 
The validity of the assumption is discussed later in this subsection.  
We replace the Sun by 2MASS~J2224--0158 as the central object. 
We take the mass, 
$M_{\star}=0.05M_{\odot}$, as a typical brown dwarf mass, where $M_{\odot}$ is the solar mass. 
We adopt the parameters of the non-heating best-fit model, $\log g=5.0$ and $T_{\rm eff}=1700$~K. 
The stellar radius is derived as $R_{\star}=0.12R_{\odot}$ from $M_{\star}$ and $\log g$.
We set the inner boundary ($r=R_{\star}$) of the simulation at the top of the surface convection zone located at the position with the gas pressure $=10^{7.08}$ dyn cm$^{-2}$ from our model atmosphere. 

We set up an open magnetic flux tube which is similar to those on the Sun.
Recent HINODE observations show that open magnetic flux tubes in coronal holes 
are anchored at very strong magnetic field regions with $\sim$ kilo-Gauss 
\citep{Tsuneta_2008}, which is nearly equipartition to the ambient gas pressure. These 
flux tubes open quite rapidly and the average field strength is reduced to 
an order of 1-10 G in the corona \citep{Ito_2010}.  
In the present simulation for a brown dwarf, we adopt 
similar properties for our underlying flux tube, namely a super-radially open 
flux tube emanating from an equipartition kG patch. 

We inject velocity perturbations at the inner boundary. 
In particular, transverse fluctuations with respect to the radial magnetic field excite Alfv\'{e}n waves travelling upwardly. 
We adopt an amplitude of 20 \% of the sound speed at the surface with a 
wide-band spectrum in proportion to the inverse of frequency ranging from 
period of 5 to 250 seconds, by referring to HINODE observation of the solar 
surface \citep{Matsumoto_2010}; 
the spectrum is logarithmically centered at a period of 
30 - 40 seconds, which can be scaled with $H/c_s\sim c_s/g \sim 1/10$  
of the solar value ($=5$ minutes oscillation), where $H$ and $c_s$ are the 
pressure scale height and the sound speed.

\begin{figure}
\begin{center}
\includegraphics[width=95mm]{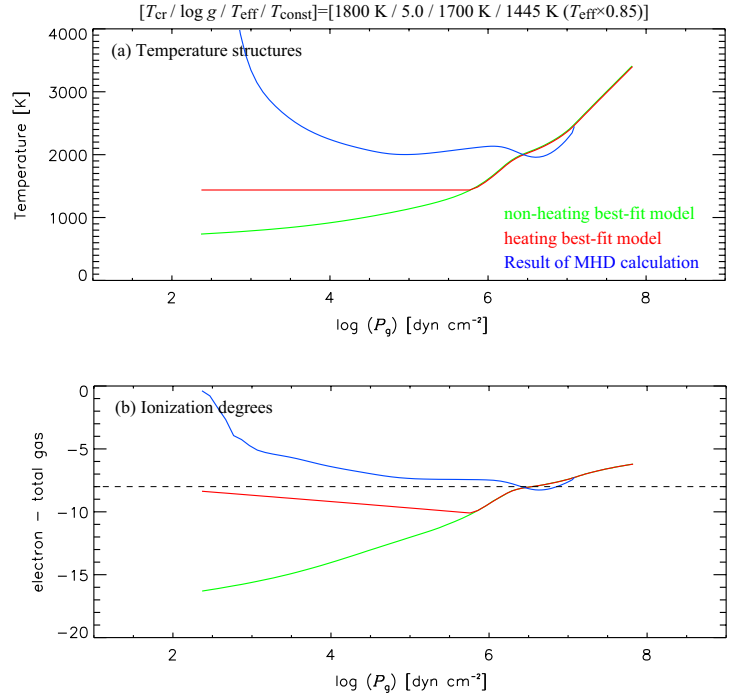}
\end{center}
\caption{Comparison of (a) temperature structures and (b) ionization degrees of the non-heating best-fit model ({\Tcr}/{\logg}/{\Teff}) = (1800K/5.0/1700K) of 2MASS~J2224--0158 ({\it green}), the heating best-fit model, and the MHD simulation ({\it blue}). }
\label{fig4}
\end{figure}

Figure~\ref{fig4} compares the temperature structures versus total gas pressure; the temperature derived from the simulation is averaged 
over sufficiently long time compared to the typical timescale of the wave propagation. 
The numerical simulation (blue line) shows that the temperature is nearly constant $\approx 2000$ K from $p=10^7$ to $10^4$ dyn cm$^{-2}$, 
and rapidly increases in $p \la 10^3$ dyn cm$^{-2}$.  The temperature actually reaches several hundred thousand K by the heating as a result 
of the dissipation of Alfv\'{e}n waves in the upper region. 
This case might be an extreme one because we are assuming the ideal MHD approximation and the more or less large velocity perturbation at the inner boundary. If the ideal MHD approximation is not satisfied, the 
amplitude of generated waves will be smaller because of magnetic diffusion 
\citep{Mohanty_2002}. Injecting smaller perturbations, 
the numerical simulation would give lower temperature, approaching to that of the simple model with $T_{\rm const}$ (red line). 
Although our treatment of the heating models with $T_{\rm const}$ is quite a simple one, we expect that it could give meaningful results.

We here examine the validity of our assumption of the ideal MHD approximation 
for the numerical simulation. 
The evolution of magnetic field is determined by an induction equation, 
\begin{equation}
\frac{\partial\mbf{B}}{\partial t} = \mbf{\nabla} \times [\mbf{v\times B} 
- \eta (\mbf{\nabla \times B}) ],  
\label{eq:ind}
\end{equation}
where $\eta$ is resistivity. Although in our simulations $\eta=0$ is assumed, 
if the second term on the right hand side dominates the first term, magnetic 
field is not well coupled to ambient gas and diffuses away. 
In the situation of a brown dwarf atmosphere, 
the collision between electrons and neutrals, which corresponds to the 
``decoupled diffusion'' term in \citet{Mohanty_2002}, is the dominant 
mechanism that accounts for the resistivity. 
This can be expressed as 
\begin{equation}
\eta \approx 200 \frac{\sqrt{T}}{x_{\rm e}} ({\rm cm^2s^{-1}})
\label{eq:restv}
\end{equation} 
\citep{Blaes_1994, Inutsuka_2005}, where $x_e$ is an ionization degree and temperature, 
$T$, is in units of Kelvin. By using this 
expression, we estimate whether the magnetic diffusion becomes significant 
or not. 

We introduce a magnetic Reynolds number, 
\begin{equation}
R_{\rm m} = v L / \eta, 
\end{equation}
which is a nondimensional variable that measures the frozen-in condition of 
magnetic field; $R_{\rm m}$ is the ratio of the first term to the second term on 
the right-hand side of Equation (\ref{eq:ind}) by replacing the rotation derivative 
$(\mbf{\nabla \times})$ via a simple division by a typical length, $L$.
The ideal MHD condition is valid if $R_{\rm m}$ is significantly larger than unity.
As a representative quantity for $L$, we can reasonably use the wavelength 
of the typical Alfv\'{e}n wave we are injecting:
\begin{equation}
L \sim v_{\rm A} \tau \sim c_{\rm s}\tau = 120\; {\rm km}\left(\frac{c_{\rm s}}
{3\;{\rm km\; s^{-1}}}\right)\left(\frac{\tau}{40\;{\rm s}}\right), 
\label{eq:wvlg}
\end{equation} 
where $\tau$ is the wave period normalized by the logarithmically centered 
value, 40 s, and $v_{\rm A}$ is the Alfv\'{e}n velocity, 
which is comparable to the sound speed, $c_{\rm s}$, because we consider 
the equipartition magnetic flux tube. Here the normalization of 
$c_{\rm s}=3$ km s$^{-1}$ corresponds to $T=1500$ K.  
Using Equations (\ref{eq:restv}) \& (\ref{eq:wvlg}), we can estimate 
\begin{equation}
R_{\rm m} = 1 \left(\frac{v}{0.6\;{\rm km\;s^{-1}}}\right)
\left(\frac{\tau}{40\;{\rm s}}\right)\left(\frac{x_{\rm e}}{10^{-8}}\right)
\label{eq:Reyval}
\end{equation}
where we normalize $v$ by the velocity amplitude ($=0.2 c_{\rm s}$ ) of the 
injected Alfv\'{e}n waves near the inner boundary. This is a conservative 
estimate because the amplitude of the Alfv\'{e}n waves is amplified as they
propagate through the density decreasing atmosphere. 
Note also that the dependence on temperature (Equations \ref{eq:restv} 
\& \ref{eq:wvlg}) is canceled out because $c_{\rm s} \propto \sqrt{T}$. 
Equation (\ref{eq:Reyval}) shows that the magnetic diffusion is not so 
significant for low-frequency ($\tau=40$ s) Alfv\'{e}n waves, even though 
the ionization degree is not so high, $x_{\rm e} > 10^{-8}$.

The ionization degrees of the non-heating best-fit model, the heating best-fit model, and the MHD simulation of 2MASS~J2224--0158 as a function of total gas pressure are shown in Figure~\ref{fig4}~(b). 
The ionization degree of the non-heating model monotonically decreases with elevating height (decreasing total gas pressure).
On the other hand, those of the heating model and the MHD simulation tend to increase, 
because these model atmospheres have higher temperature and lower density than non-heating model.
The ionization degree of the heating model decreases with elevating height (the same as the non-heating model) 
until reaching the region with $T=${\Tconst}, and then increases toward the upper region.
The ionization degree resulting from the MHD simulation is larger 
than those of the other two cases and exceeds 10$^{-8}$ in the almost entire 
region except for the location around log$P_g \sim 6.5$.
Therefore, the ideal MHD approximation is marginally acceptable for 
this case.
%
In more elaborated studies, we should solve resistive MHD equations by using 
derived an ionization degree in a self-consistent manner.

In the above estimate, we only take into account thermal ionization. 
However, additional ionization processes are supposed to work 
in the atmosphere of brown dwarfs. Helling and her collaborators have 
proposed various ionization mechanisms, e.g. collision between 
charged dust grains \citep{2011ApJ...727....4H, 2011ApJ...737...38H}, 
inter-grain electrical discharge \citep{Helling_2013}, ionization by external 
cosmic rays \citep{Rimmer_2013}, and Alfv\'{e}n ionization \citep{Stark_2013}. 
If these processes actually work, the ionization degree will be larger than 
the above estimate, leading to better coupling between gas and magnetic field.

Observations show that the ratio of X-ray and H${\alpha}$ luminosities to 
bolometric luminosity appears to decrease with later spectral type, while 
the ratio of radio luminosity to bolometric luminosity 
increase with later spectral type \citealt{Berger_2010, Hallinan_2007}
If we take into account magnetic diffusivity in our MHD simulations, we expect 
that the tendency of X-ray and H${\alpha}$ will be interpreted at least in a 
qualitative sense by decreasing $x_{\rm e}$ with decreasing atmospheric 
temperature \citep{Mohanty_2002}. In contrast, the radio luminosity is supposed 
to be emitted from non-thermal electrons, which is beyond the scope of our 
MHD simulations that handle the thermal component only.

\subsection{Dust effects}
For the ``unsuccessful'' four objects in Figure~\ref{fig2} excluding 2MASS~J0036+1821, 
the flux levels around 4.0~$\mu$m differ between the observations and the model spectra.
In this study, we focus the only upper temperature structure; i.e., we do not modify inner atmospheric structure affected by dust. 
However, the mid-L dwarfs are supposed to be most affected by the dust in their photosphere, 
thus their atmospheres should be complicated. 
Our study shows the flux level of the 4.0~$\mu$m region is affected 
by dust volume (Sorahana et al. 2013 submitted to ApJ). 
Therefor we may need to consider some additional dust effects along with {\Tcr} in UCM, for example, changing abundances, distributions, providing size distribution, and adding other dust species. 
\citet{2008MNRAS.391.1854H} compared five models of brown dwarf atmospheres. 
The other models constructed by Marley, Ackerman \& Lodders, Allard \& Homeier and Helling \& Woitke consider vertical mixing efficiency. 
\citet{Yamamura_2010} showed that for L dwarfs a vertical mixing in the surface of the photosphere does not affects to molecular abundances in that region, thus spectral features also does not change. 
Grain size distributions calculated by comparing between
time-scales for mixing due to convective overshooting and condensation and gravitational settling 
are not implemented in the UCM. 
We also need to consider additional effects such as hydrodynamic processes including meteorological aspect.

\section{Conclusions}

To solve the discrepancy between observed and model spectra between 1.0 and 5.0~$\mu$m, 
we consider the additional effects concerning chromospheric activity, coronae, and flares which 
possibly affect the temperature structure in an upper atmosphere. 
First, we carry out the model fittings to the only SpeX/CGS4 spectra to pin down the temperature structures in the deeper atmospheres. 
After that, we change the upper thermal structure in the derived model photosphere
with a temperature floor, $T_{\rm const}$, to take into account the effect of the chromosphere. 
Then we compare the heating model spectra with the observed spectra for eight brown dwarfs taken by {\AKARI}.
We validate that the spectrum of 2.5--5.0~$\mu$m reflects the structure of the upper photosphere;
in particular, the 3.3~$\mu$m {\CHf}, 2.7~$\mu$m {\HtO} and 4.6~$\mu$m CO bands are sensitive to the thermal structure of the upper photosphere region. 
From the comparison between the observed and heating model spectra, 
we find that three objects with relatively strong $H{\alpha}$ emission are consistently explained 
by the model spectra with $T_{\rm const}$ owing to the additional heating. 
We carry out the MHD simulation for a brown dwarf atmosphere by 
extending the simulation code originally developed for the Sun.  
The numerical simulation indeed shows that the temperature is kept nearly constant in the atmosphere and eventually increases in the upper region. 
Other four mid-L objects cannot be explained by our current heating model, especially the flux levels around 4.0~$\mu$m.
We may need to reconsider inner atmospheric structure with additional dust effects 
or some hydrodynamic processes.

\section*{Acknowledgments}

We thank the anonymous referee for providing critical comments and constructive suggestions to our article .
This research is based on observations with {\AKARI}, a JAXA project with the participation of ESA.  
Dr. Adam Burgasser, Dr. Michael Cushing, and Dr. Dagny Looper kindly provide us the spectral data and warm encouragement. 
I have greatly benefited from Prof. Ansgar Reiners. 
We thank to Prof. Takashi Tsuji for his kind permission to access the UCM and helpful suggestions. 
We also thank Dr. Takuma Matsumoto for useful discussions throughout this research, 
and thank Dr. Jennifer Stone for her careful checking our paper.
This work is  supported in part by Grants-in-Aid for 
Scientific Research from the MEXT of Japan, 22864006. (PI: T.K.S.)

\bibliography{sorahana20140308}

\begin{thebibliography}{}

\bibitem[\protect\citeauthoryear{{Ackerman} \& {Marley}}{{Ackerman} \&
  {Marley}}{2001}]{Ackerman_2001}
{Ackerman} A.~S.,  {Marley} M.~S.,  2001, ApJ, 556, 872

\bibitem[\protect\citeauthoryear{{Allard}, {Hauschildt}, {Alexander}, {Tamanai}
  \& {Schweitzer}}{{Allard} et~al.}{2001}]{Allard_2001}
{Allard} F.,  {Hauschildt} P.~H.,  {Alexander} D.~R.,  {Tamanai} A.,
  {Schweitzer} A.,  2001, ApJ, 556, 357

\bibitem[\protect\citeauthoryear{{Allard}, {Allard}, {Hauschildt}, {Kielkopf}
  \& {Machin}}{{Allard} et~al.}{2003}]{Allard_2003}
{Allard} N.~F.,  {Allard} F.,  {Hauschildt} P.~H.,  {Kielkopf} J.~F.,
  {Machin} L.,  2003, A\&A, 411, L473

\bibitem[\protect\citeauthoryear{{Anderson} \& {Athay}}{{Anderson} \&
  {Athay}}{1989}]{Anderson_1989}
{Anderson} L.~S.,  {Athay} R.~G.,  1989, ApJ, 336, 1089

\bibitem[\protect\citeauthoryear{{Baraffe}, {Chabrier}, {Allard} \&
  {Hauschildt}}{{Baraffe} et~al.}{2002}]{Baraffe_2002}
{Baraffe} I.,  {Chabrier} G.,  {Allard} F.,    {Hauschildt} P.~H.,  2002, A\&A,
  382, 563

\bibitem[\protect\citeauthoryear{{Berger}, {Basri}, {Fleming}, {Giampapa},
  {Gizis}, {Liebert}, {Mart{\'{\i}}n}, {Phan-Bao} \& {Rutledge}}{{Berger}
  et~al.}{2010}]{Berger_2010}
{Berger} E.,  {Basri} G.,  {Fleming} T.~A.,  {Giampapa} M.~S.,  {Gizis} J.~E.,
  {Liebert} J.,  {Mart{\'{\i}}n} E.,  {Phan-Bao} N.,    {Rutledge} R.~E.,
  2010, ApJ, 709, 332

\bibitem[\protect\citeauthoryear{{Bingham}, {Cairns} \& {Kellett}}{{Bingham}
  et~al.}{2001}]{Bingham_2001a}
{Bingham} R.,  {Cairns} R.~A.,    {Kellett} B.~J.,  2001, A\&A, 370, 1000

\bibitem[\protect\citeauthoryear{{Blaes} \& {Balbus}}{{Blaes} \&
  {Balbus}}{1994}]{Blaes_1994}
{Blaes} O.~M.,  {Balbus} S.~A.,  1994, ApJ, 421, 163

\bibitem[\protect\citeauthoryear{{Burgasser}}{{Burgasser}}{2007}]{Burgasser_2007}
{Burgasser} A.~J.,  2007, ApJ, 659, 655

\bibitem[\protect\citeauthoryear{{Burgasser}, {Cruz}, {Cushing}, {Gelino},
  {Looper} et~al.,}{{Burgasser} et~al.}{2010}]{Burgasser_2010}
{Burgasser} A.~J.,  {Cruz} K.~L.,  {Cushing} M.,  {Gelino} C.~R.,  {Looper}
  D.~L.,    et~al., 2010, ApJ, 710, 1142

\bibitem[\protect\citeauthoryear{{Burgasser}, {Geballe}, {Leggett},
  {Kirkpatrick} \& {Golimowski}}{{Burgasser} et~al.}{2006}]{Burgasser_2006}
{Burgasser} A.~J.,  {Geballe} T.~R.,  {Leggett} S.~K.,  {Kirkpatrick} J.~D.,
  {Golimowski} D.~A.,  2006, ApJ, 637, 1067

\bibitem[\protect\citeauthoryear{{Burgasser}, {Liu}, {Ireland}, {Cruz} \&
  {Dupuy}}{{Burgasser} et~al.}{2008}]{Burgasser_2008}
{Burgasser} A.~J.,  {Liu} M.~C.,  {Ireland} M.~J.,  {Cruz} K.~L.,    {Dupuy}
  T.~J.,  2008, ApJ, 681, 579

\bibitem[\protect\citeauthoryear{{Burgasser}, {McElwain}, {Kirkpatrick},
  {Cruz}, {Tinney} \& {Reid}}{{Burgasser} et~al.}{2004}]{Burgasser_2004}
{Burgasser} A.~J.,  {McElwain} M.~W.,  {Kirkpatrick} J.~D.,  {Cruz} K.~L.,
  {Tinney} C.~G.,    {Reid} I.~N.,  2004, AJ, 127, 2856

\bibitem[\protect\citeauthoryear{{Burrows}, {Hubbard}, {Lunine} \&
  {Liebert}}{{Burrows} et~al.}{2001}]{Burrows_2001}
{Burrows} A.,  {Hubbard} W.~B.,  {Lunine} J.~I.,    {Liebert} J.,  2001,
  Reviews of Modern Physics, 73, 719

\bibitem[\protect\citeauthoryear{{Choudhuri}, {Schussler} \&
  {Dikpati}}{{Choudhuri} et~al.}{1995}]{Choudhuri_1995}
{Choudhuri} A.~R.,  {Schussler} M.,    {Dikpati} M.,  1995, A\&A, 303, L29

\bibitem[\protect\citeauthoryear{{Cooper}, {Sudarsky}, {Milsom}, {Lunine} \&
  {Burrows}}{{Cooper} et~al.}{2003}]{Cooper_2003}
{Cooper} C.~S.,  {Sudarsky} D.,  {Milsom} J.~A.,  {Lunine} J.~I.,    {Burrows}
  A.,  2003, ApJ, 586, 1320

\bibitem[\protect\citeauthoryear{{Cushing}, {Marley}, {Saumon}, {Kelly},
  {Vacca}, {Rayner}, {Freedman}, {Lodders} \& {Roellig}}{{Cushing}
  et~al.}{2008}]{Cushing_2008}
{Cushing} M.~C.,  {Marley} M.~S.,  {Saumon} D.,  {Kelly} B.~C.,  {Vacca} W.~D.,
   {Rayner} J.~T.,  {Freedman} R.~S.,  {Lodders} K.,    {Roellig} T.~L.,  2008,
  ApJ, 678, 1372

\bibitem[\protect\citeauthoryear{{Cushing}, {Roellig}, {Marley}, {Saumon},
  {Leggett}, {Kirkpatrick}, {Wilson}, {Sloan}, {Mainzer}, {Van Cleve} \&
  {Houck}}{{Cushing} et~al.}{2006}]{Cushing_2006}
{Cushing} M.~C.,  {Roellig} T.~L.,  {Marley} M.~S.,  {Saumon} D.,  {Leggett}
  S.~K.,  {Kirkpatrick} J.~D.,  {Wilson} J.~C.,  {Sloan} G.~C.,  {Mainzer}
  A.~K.,  {Van Cleve} J.~E.,    {Houck} J.~R.,  2006, ApJ, 648, 614

\bibitem[\protect\citeauthoryear{{Cushing}, {Vacca} \& {Rayner}}{{Cushing}
  et~al.}{2004}]{Cushing_2004}
{Cushing} M.~C.,  {Vacca} W.~D.,    {Rayner} J.~T.,  2004, PASP, 116, 362

\bibitem[\protect\citeauthoryear{{Geballe}, {Knapp}, {Leggett}, {Fan},
  {Golimowski}, {Anderson}, {Brinkmann}, {Csabai}, {Gunn}, {Hawley}
  et~al.,}{{Geballe} et~al.}{2002}]{Geballe_2002}
{Geballe} T.~R.,  {Knapp} G.~R.,  {Leggett} S.~K.,  {Fan} X.,  {Golimowski}
  D.~A.,  {Anderson} S.,  {Brinkmann} J.,  {Csabai} I.,  {Gunn} J.~E.,
  {Hawley} S.~L.,    et~al., 2002, ApJ, 564, 466

\bibitem[\protect\citeauthoryear{{Gizis}, {Monet}, {Reid}, {Kirkpatrick},
  {Liebert} \& {Williams}}{{Gizis} et~al.}{2000}]{Gizis_2000}
{Gizis} J.~E.,  {Monet} D.~G.,  {Reid} I.~N.,  {Kirkpatrick} J.~D.,  {Liebert}
  J.,    {Williams} R.~J.,  2000, AJ, 120, 1085

\bibitem[\protect\citeauthoryear{{Hallinan}, {Antonova}, {Doyle}, {Bourke},
  {Lane} \& {Golden}}{{Hallinan} et~al.}{2008}]{Hallinan_2008}
{Hallinan} G.,  {Antonova} A.,  {Doyle} J.~G.,  {Bourke} S.,  {Lane} C.,
  {Golden} A.,  2008, ApJ, 684, 644

\bibitem[\protect\citeauthoryear{{Hallinan}, {Bourke}, {Lane}, {Antonova},
  {Zavala}, {Brisken}, {Boyle}, {Vrba}, {Doyle} \& {Golden}}{{Hallinan}
  et~al.}{2007}]{Hallinan_2007}
{Hallinan} G.,  {Bourke} S.,  {Lane} C.,  {Antonova} A.,  {Zavala} R.~T.,
  {Brisken} W.~F.,  {Boyle} R.~P.,  {Vrba} F.~J.,  {Doyle} J.~G.,    {Golden}
  A.,  2007, ApJL, 663, L25

\bibitem[\protect\citeauthoryear{{Helling}, {Ackerman}, {Allard}, {Dehn},
  {Hauschildt}, {Homeier}, {Lodders}, {Marley}, {Rietmeijer}, {Tsuji} \&
  {Woitke}}{{Helling} et~al.}{2008}]{2008MNRAS.391.1854H}
{Helling} C.,  {Ackerman} A.,  {Allard} F.,  {Dehn} M.,  {Hauschildt} P.,
  {Homeier} D.,  {Lodders} K.,  {Marley} M.,  {Rietmeijer} F.,  {Tsuji} T.,
  {Woitke} P.,  2008, MNRAS, 391, 1854

\bibitem[\protect\citeauthoryear{{Helling}, {Jardine} \& {Mokler}}{{Helling}
  et~al.}{2011}]{2011ApJ...737...38H}
{Helling} C.,  {Jardine} M.,    {Mokler} F.,  2011, ApJ, 737, 38

\bibitem[\protect\citeauthoryear{{Helling}, {Jardine}, {Stark} \&
  {Diver}}{{Helling} et~al.}{2013}]{Helling_2013}
{Helling} C.,  {Jardine} M.,  {Stark} C.,    {Diver} D.,  2013, ApJ, 767, 136

\bibitem[\protect\citeauthoryear{{Helling}, {Jardine}, {Witte} \&
  {Diver}}{{Helling} et~al.}{2011}]{2011ApJ...727....4H}
{Helling} C.,  {Jardine} M.,  {Witte} S.,    {Diver} D.~A.,  2011, ApJ, 727, 4

\bibitem[\protect\citeauthoryear{{Helling}, {Oevermann}, {L{\"u}ttke}, {Klein}
  \& {Sedlmayr}}{{Helling} et~al.}{2001}]{Helling_2001}
{Helling} C.,  {Oevermann} M.,  {L{\"u}ttke} M.~J.~H.,  {Klein} R.,
  {Sedlmayr} E.,  2001, A\&A, 376, 194

\bibitem[\protect\citeauthoryear{{Helling} \& {Woitke}}{{Helling} \&
  {Woitke}}{2006}]{Helling_2006}
{Helling} C.,  {Woitke} P.,  2006, A\&A, 455, 325

\bibitem[\protect\citeauthoryear{{Helling}, {Woitke} \& {Thi}}{{Helling}
  et~al.}{2008}]{2008A&A...485..547H}
{Helling} C.,  {Woitke} P.,    {Thi} W.-F.,  2008, A\&A, 485, 547

\bibitem[\protect\citeauthoryear{{Hotta}, {Rempel}, {Yokoyama}, {Iida} \&
  {Fan}}{{Hotta} et~al.}{2012}]{Hotta_2012}
{Hotta} H.,  {Rempel} M.,  {Yokoyama} T.,  {Iida} Y.,    {Fan} Y.,  2012, A\&A,
  539, A30

\bibitem[\protect\citeauthoryear{{Inutsuka} \& {Sano}}{{Inutsuka} \&
  {Sano}}{2005}]{Inutsuka_2005}
{Inutsuka} S.-i.,  {Sano} T.,  2005, ApJL, 628, L155

\bibitem[\protect\citeauthoryear{{Ito}, {Tsuneta}, {Shiota}, {Tokumaru} \&
  {Fujiki}}{{Ito} et~al.}{2010}]{Ito_2010}
{Ito} H.,  {Tsuneta} S.,  {Shiota} D.,  {Tokumaru} M.,    {Fujiki} K.,  2010,
  ApJ, 719, 131

\bibitem[\protect\citeauthoryear{{Kellett}, {Bingham}, {Cairns} \&
  {Tsikoudi}}{{Kellett} et~al.}{2002}]{Kellett_2002}
{Kellett} B.~J.,  {Bingham} R.,  {Cairns} R.~A.,    {Tsikoudi} V.,  2002,
  MNRAS, 329, 102

\bibitem[\protect\citeauthoryear{{Kirkpatrick}, {Reid}, {Liebert}, {Gizis},
  {Burgasser}, {Monet}, {Dahn}, {Nelson} \& {Williams}}{{Kirkpatrick}
  et~al.}{2000}]{Kirkpatrick_2000}
{Kirkpatrick} J.~D.,  {Reid} I.~N.,  {Liebert} J.,  {Gizis} J.~E.,  {Burgasser}
  A.~J.,  {Monet} D.~G.,  {Dahn} C.~C.,  {Nelson} B.,    {Williams} R.~J.,
  2000, AJ, 120, 447

\bibitem[\protect\citeauthoryear{{Marley}, {Seager}, {Saumon}, {Lodders},
  {Ackerman} et~al.,}{{Marley} et~al.}{2002}]{Marley_2002}
{Marley} M.~S.,  {Seager} S.,  {Saumon} D.,  {Lodders} K.,  {Ackerman} A.~S.,
   et~al., 2002, ApJ, 568, 335

\bibitem[\protect\citeauthoryear{{Matsumoto} \& {Kitai}}{{Matsumoto} \&
  {Kitai}}{2010}]{Matsumoto_2010}
{Matsumoto} T.,  {Kitai} R.,  2010, ApJL, 716, L19

\bibitem[\protect\citeauthoryear{{Matsumoto} \& {Suzuki}}{{Matsumoto} \&
  {Suzuki}}{2012}]{Matsumoto_2012}
{Matsumoto} T.,  {Suzuki} T.~K.,  2012, ApJ, 749, 8

\bibitem[\protect\citeauthoryear{{McLean}, {Berger} \& {Reiners}}{{McLean}
  et~al.}{2012}]{McLean_2012}
{McLean} M.,  {Berger} E.,    {Reiners} A.,  2012, ApJ, 746, 23

\bibitem[\protect\citeauthoryear{{Mohanty}, {Baraffe} \& {Chabrier}}{{Mohanty}
  et~al.}{2007}]{Mohanty_2007}
{Mohanty} S.,  {Baraffe} I.,    {Chabrier} G.,  2007, in {Kupka} F.,
  {Roxburgh} I.,   {Chan} K.~L.,  eds, IAU Symposium Vol.~239 of IAU Symposium,
  {Convection in brown dwarfs}.
pp 197--204

\bibitem[\protect\citeauthoryear{{Mohanty} \& {Basri}}{{Mohanty} \&
  {Basri}}{2003}]{Mohanty_2003}
{Mohanty} S.,  {Basri} G.,  2003, ApJ, 583, 451

\bibitem[\protect\citeauthoryear{{Mohanty}, {Basri}, {Shu}, {Allard} \&
  {Chabrier}}{{Mohanty} et~al.}{2002}]{Mohanty_2002}
{Mohanty} S.,  {Basri} G.,  {Shu} F.,  {Allard} F.,    {Chabrier} G.,  2002,
  ApJ, 571, 469

\bibitem[\protect\citeauthoryear{{Nakajima}, {Oppenheimer}, {Kulkarni},
  {Golimowski}, {Matthews} \& {Durrance}}{{Nakajima}
  et~al.}{1995}]{Nakajima_1995}
{Nakajima} T.,  {Oppenheimer} B.~R.,  {Kulkarni} S.~R.,  {Golimowski} D.~A.,
  {Matthews} K.,    {Durrance} S.~T.,  1995, Nature, 378, 463

\bibitem[\protect\citeauthoryear{{Nakajima}, {Tsuji} \&
  {Yanagisawa}}{{Nakajima} et~al.}{2001}]{Nakajima_2001}
{Nakajima} T.,  {Tsuji} T.,    {Yanagisawa} K.,  2001, ApJ, 561, L119

\bibitem[\protect\citeauthoryear{{Reiners} \& {Basri}}{{Reiners} \&
  {Basri}}{2008}]{Reiners_2008}
{Reiners} A.,  {Basri} G.,  2008, ApJ, 684, 1390

\bibitem[\protect\citeauthoryear{{Rimmer} \& {Helling}}{{Rimmer} \&
  {Helling}}{2013}]{Rimmer_2013}
{Rimmer} P.~B.,  {Helling} C.,  2013, ApJ, 774, 108

\bibitem[\protect\citeauthoryear{{Schmidt}, {Cruz}, {Bongiorno}, {Liebert} \&
  {Reid}}{{Schmidt} et~al.}{2007}]{Schmidt_2007}
{Schmidt} S.~J.,  {Cruz} K.~L.,  {Bongiorno} B.~J.,  {Liebert} J.,    {Reid}
  I.~N.,  2007, AJ, 133, 2258

\bibitem[\protect\citeauthoryear{{Sorahana} \& {Yamamura}}{{Sorahana} \&
  {Yamamura}}{2012}]{Sorahana_2012}
{Sorahana} S.,  {Yamamura} I.,  2012, ApJ, 760, 151

\bibitem[\protect\citeauthoryear{{Stark}, {Helling}, {Diver} \&
  {Rimmer}}{{Stark} et~al.}{2013}]{Stark_2013}
{Stark} C.~R.,  {Helling} C.,  {Diver} D.~A.,    {Rimmer} P.~B.,  2013, ApJ,
  776, 11

\bibitem[\protect\citeauthoryear{{Stelzer}, {Micela}, {Flaccomio},
  {Neuh{\"a}user} \& {Jayawardhana}}{{Stelzer} et~al.}{2006}]{Stelzer_2006}
{Stelzer} B.,  {Micela} G.,  {Flaccomio} E.,  {Neuh{\"a}user} R.,
  {Jayawardhana} R.,  2006, A\&A, 448, 293

\bibitem[\protect\citeauthoryear{{Suzuki}}{{Suzuki}}{2007}]{Suzuki_2007}
{Suzuki} T.~K.,  2007, ApJ, 659, 1592

\bibitem[\protect\citeauthoryear{{Suzuki}, {Imada}, {Kataoka}, {Kato},
  {Matsumoto}, {Miyahara} \& {Tsuneta}}{{Suzuki} et~al.}{2013}]{Suzuki_2013}
{Suzuki} T.~K.,  {Imada} S.,  {Kataoka} R.,  {Kato} Y.,  {Matsumoto} T.,
  {Miyahara} H.,    {Tsuneta} S.,  2013, ArXiv e-prints, PASJ in press

\bibitem[\protect\citeauthoryear{{Suzuki} \& {Inutsuka}}{{Suzuki} \&
  {Inutsuka}}{2005}]{Suzuki_2005}
{Suzuki} T.~K.,  {Inutsuka} S.-i.,  2005, ApJL, 632, L49

\bibitem[\protect\citeauthoryear{{Suzuki} \& {Inutsuka}}{{Suzuki} \&
  {Inutsuka}}{2006}]{Suzuki_2006}
{Suzuki} T.~K.,  {Inutsuka} S.-I.,  2006, Journal of Geophysical Research
  (Space Physics), 111, 6101

\bibitem[\protect\citeauthoryear{{Tanaka}, {Suzuki} \& {Inutsuka}}{{Tanaka}
  et~al.}{2013}]{Tanaka_2013}
{Tanaka} Y.~A.,  {Suzuki} T.~K.,    {Inutsuka} S.-i.,  2013, ArXiv e-prints

\bibitem[\protect\citeauthoryear{{Tsuboi}, {Maeda}, {Feigelson}, {Garmire},
  {Chartas}, {Mori} \& {Pravdo}}{{Tsuboi} et~al.}{2003}]{Tsuboi_2003}
{Tsuboi} Y.,  {Maeda} Y.,  {Feigelson} E.~D.,  {Garmire} G.~P.,  {Chartas} G.,
  {Mori} K.,    {Pravdo} S.~H.,  2003, ApJL, 587, L51

\bibitem[\protect\citeauthoryear{{Tsuji}}{{Tsuji}}{2002}]{Tsuji_2002}
{Tsuji} T.,  2002, ApJ, 575, 264

\bibitem[\protect\citeauthoryear{{Tsuji}}{{Tsuji}}{2005}]{Tsuji_2005}
{Tsuji} T.,  2005, ApJ, 621, 1033

\bibitem[\protect\citeauthoryear{{Tsuji}, {Ohnaka} \& {Aoki}}{{Tsuji}
  et~al.}{1996}]{Tsuji_1996a}
{Tsuji} T.,  {Ohnaka} K.,    {Aoki} W.,  1996, A\&A, 305, L1

\bibitem[\protect\citeauthoryear{{Tsuji}, {Ohnaka}, {Aoki} \&
  {Nakajima}}{{Tsuji} et~al.}{1996}]{Tsuji_1996b}
{Tsuji} T.,  {Ohnaka} K.,  {Aoki} W.,    {Nakajima} T.,  1996, A\&A, 308, L29

\bibitem[\protect\citeauthoryear{{Tsuneta}, {Ichimoto}, {Katsukawa}, {Lites},
  {Matsuzaki}, {Nagata}, {Orozco Su{\'a}rez}, {Shimizu}, {Shimojo}, {Shine},
  {Suematsu}, {Suzuki}, {Tarbell} \& {Title}}{{Tsuneta}
  et~al.}{2008}]{Tsuneta_2008}
{Tsuneta} S.,  {Ichimoto} K.,  {Katsukawa} Y.,  {Lites} B.~W.,  {Matsuzaki} K.,
   {Nagata} S.,  {Orozco Su{\'a}rez} D.,  {Shimizu} T.,  {Shimojo} M.,  {Shine}
  R.~A.,  {Suematsu} Y.,  {Suzuki} T.~K.,  {Tarbell} T.~D.,    {Title} A.~M.,
  2008, ApJ, 688, 1374

\bibitem[\protect\citeauthoryear{{Woitke} \& {Helling}}{{Woitke} \&
  {Helling}}{2003}]{Woitke_2003}
{Woitke} P.,  {Helling} C.,  2003, A\&A, 399, 297

\bibitem[\protect\citeauthoryear{{Woitke} \& {Helling}}{{Woitke} \&
  {Helling}}{2004}]{Woitke_2004}
{Woitke} P.,  {Helling} C.,  2004, A\&A, 414, 335

\bibitem[\protect\citeauthoryear{{Yamamura}, {Tsuji} \&
  {Tanab{\'e}}}{{Yamamura} et~al.}{2010}]{Yamamura_2010}
{Yamamura} I.,  {Tsuji} T.,    {Tanab{\'e}} T.,  2010, ApJ, 722, 682

\end{thebibliography}


\label{lastpage}

\end{document}